\documentclass[12pt]{article}
\usepackage{times}
\usepackage{geometry}
\usepackage[utf8]{inputenc}
\usepackage{enumitem,amssymb}
\usepackage{ragged2e}
\newlist{thematic}{itemize}{8}
\setlist[thematic]{label=$\square$}
\usepackage{pifont}
\newcommand{\cmark}{\ding{51}}%
\newcommand{\done}{\rlap{$\square$}{\raisebox{2pt}{\large\hspace{1pt}\cmark}}%
\hspace{-2.5pt}}

\usepackage{pslatex}
\usepackage[square,comma,numbers]{natbib}
\usepackage{color}
\usepackage{graphicx}
\usepackage{upgreek}
\usepackage{amsmath, amsthm, amssymb}
\usepackage{sidecap}
\sidecaptionvpos{figure}{c}

\usepackage{sectsty}
\sectionfont{\fontsize{12}{15}\selectfont}

\begin{document}
\raggedright
\huge
Astro 2020 Science White Paper
\bigskip
\linebreak Fundamental Cosmology in the Dark Ages 
\linebreak
with 21-cm Line Fluctuations \linebreak
\normalsize

\noindent \textbf{Thematic Areas:} \hspace*{60pt} $\square$ Planetary Systems \hspace*{10pt} $\square$ Star and Planet Formation \hspace*{20pt}\linebreak
$\square$ Formation and Evolution of Compact Objects \hspace*{31pt} $\done$ Cosmology and Fundamental Physics \linebreak
  $\square$  Stars and Stellar Evolution \hspace*{1pt} $\square$ Resolved Stellar Populations and their Environments \hspace*{40pt} \linebreak
  $\square$    Galaxy Evolution   \hspace*{45pt} $\square$             Multi-Messenger Astronomy and Astrophysics \hspace*{65pt} \linebreak
  
\textbf{Principal Author:}

Name:	Steven R. Furlanetto
 \linebreak						
Institution:  University of California Los Angeles
 \linebreak
Email: sfurlane@astro.ucla.edu
 \linebreak
Phone:  (310) 206-4127
 \linebreak
 
\textbf{Co-authors:} (names and institutions)  Judd D. Bowman (Arizona State University), Jordan Mirocha (McGill University), Jonathan C. Pober (Brown University), Jack Burns (University of Colorado Boulder), Chris L. Carilli (NRAO, Cavendish Laboratory), Julian Munoz (Harvard University), James Aguirre (University of Pennsylvania), Yacine Ali-Haimoud (New York University), Marcelo Alvarez (University of California Berkeley), Adam Beardsley (Arizona State University), George Becker (University of California Riverside), Patrick Breysse (CITA), Volker Bromm (University of Texas at Austin), Philip Bull (Queen Mary  University of London), Tzu-Ching Chang (JPL), Xuelei Chen (National Astronomical Observatory of China), Hsin Chiang (McGill University), Joanne Cohn (University of California Berkeley), Frederick Davies (University of California Santa Barbara), David DeBoer (University of California Berkeley), Joshua Dillon (University of California Berkeley), Olivier Dor\'{e} (JPL, California Institute of Technology), Cora Dvorkin (Harvard University), Anastasia Fialkov (University of Sussex), Bryna Hazelton (University of Washington), Daniel Jacobs (Arizona State University), Kirit Karkare (University of Chicago/KICP), Saul Kohn (The Vanguard Group), Leon Koopmans (Kapteyn Astronomical Institute), Ely Kovetz (Ben-Gurion University), Paul La Plante (University of Pennsylvania), Adam Lidz (University of Pennsylvania), Adrian Liu (McGill University), Yin-Zhe Ma (University of KwaZulu-Natal), Yi Mao (Tsinghua University), Kiyoshi Masui (MIT Kavli Institute for Astrophysics and Space Research), Andrew Mesinger (Scuola Normale Superiore), Steven Murray (Arizona State University), Aaron Parsons (University of California Berkeley), Benjamin Saliwanchik (Yale University), Jonathan Sievers (McGill University), Nithyanandan Thyagarajan (NRAO), Hy Trac (Carnegie Mellon University), Eli Visbal (Flatiron Institute), Matias Zaldarriaga (Institute for Advanced Study)
  \linebreak

\textbf{Abstract:} The \emph{Dark Ages} are the period between the last scattering of the cosmic microwave background and the appearance of the first luminous sources, spanning approximately $1100 < z < 30$. The only known way to measure fluctuations in this era is through the 21-cm line of neutral hydrogen. Such observations have enormous potential for cosmology, because they span a large volume while the fluctuations remain linear even on small scales. Observations of 21-cm fluctuations during this era can therefore constrain fundamental aspects of our Universe, including inflation and any exotic physics of dark matter. While the observational challenges to these low-frequency 21-cm observations are enormous, especially from the terrestrial environment, they represent an important goal for cosmology.

\pagebreak

\RaggedRight

\section{Introduction} \vspace{-12pt} 
One of the last unexplored periods of the Universe's history are the \emph{Cosmic Dark Ages}, the period between the last scattering of the cosmic microwave background (CMB) at $z \sim 1100$ and the formation of the first luminous sources, likely near $z \sim 30$ according to recent models (e.g., \cite{Crosby2013}). This period is extraordinarily difficult to study because of the lack of luminous sources. {\bf The only potential probe of this era is  absorption of the CMB by the hyperfine transition of hydrogen in the low-density intergalactic medium (IGM).} While CMB is too cold to interact with the electronic energy levels of hydrogen, it remains relatively bright in the radio, opening the possibility of studying the Dark Ages through hydrogen's 21-cm line. Further, the absence of other sources of radiation also points to one of the most compelling reasons to study it: like the CMB, it is completely unaffected by astrophysical sources and hence can be modeled cleanly.

The hyperfine 21-cm line of hydrogen has a long history in astronomy, but efforts to observe its fluctuation power spectrum in the high-$z$ Universe are still maturing. Several telescopes are targeting this transition, including PAPER, the MWA, LOFAR, and HERA, although they all focus on the lower redshift era after the first luminous sources turn on.  While the 21-cm signal is, by some measures, relatively bright compared to CMB anisotropies (with a typical brightness temperature of $> 10$~mK), the observational challenges are enormous. For the Dark Ages signal, the relevant redshifted 21-cm line frequencies are $<50$~MHz and the Galactic synchrotron foreground is orders of magnitude brighter than the signal. Moreover, Earth's ionosphere refracts and absorbs low-frequency radio waves, especially at $\nu \lesssim 30$~MHz, making the Universe at $z \gtrsim 50$ extremely challenging to observe from the ground. Finally, low radio frequencies are heavily utilized by terrestrial broadcasts and strongly affected by atmospheric events. One way to avoid the latter two challenges is with a radio array on the lunar far side, as the Moon lacks an ionosphere and blocks terrestrial interference.

Here we argue that, despite these challenges, the Dark Ages are an enormously compelling period to study. While likely beyond the capabilities of instruments available in the next decade, Dark Ages cosmology provides an important goal toward which we must push. Moreover, the Dark Ages are so sensitive to ``exotic physics" that even upper limits provided by pathfinder instruments will provide useful constraints on dark matter, primordial black holes, and more.

\begin{SCfigure}
	\includegraphics[width=9cm]{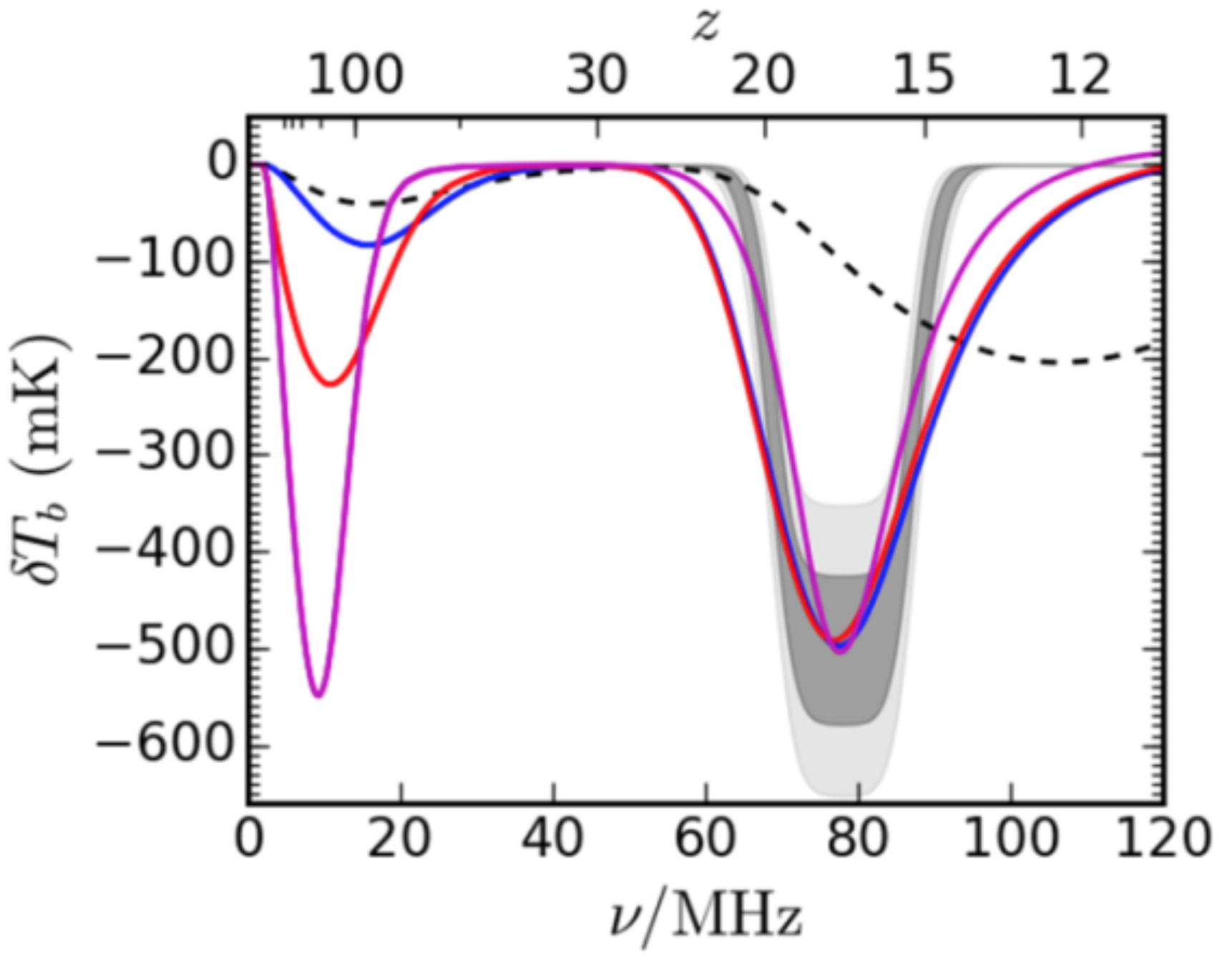} 
    \caption{{\bf The Dark Ages 21-cm absorption feature is a sensitive probe of cosmology.} The black dashed line shows the mean 21-cm brightness temperature (averaged across the sky) in a ``standard" model of cosmology. The shape at $z\gtrsim 30$ is independent of astrophysical sources. The gray contours show schematically the reported EDGES absorption signal \cite{Bowman2018}. The solid curves are phenomenological models that invoke extra cooling to match the amplitude of the EDGES signal (as in \cite{Mirocha2019}) but that also dramatically affect the Dark Ages absorption trough at $z>50$.}
    \label{fig:sample-histories}
\end{SCfigure}

\vspace{-12pt} \section{The 21-cm Background} \vspace{-12pt} 

The 21-cm brightness temperature relative to the CMB, $\delta T_{21}$, from a gas element at redshift $z$ is
\begin{equation}
\delta T_{21} \approx 60 x_{\rm HI} (1 + \delta) \left( {1 + z \over 50} \right)^{1/2} \left( {T_S - T_{\rm \gamma} \over T_S} \right) \left( {H(z)/(1+z) \over d v_\parallel / d r_\parallel} \right) \ \mbox{mK},
\label{eq:21cm-dtb}
\end{equation}
where $x_{\rm HI}$ is the neutral fraction (effectively unity throughout the Dark Ages in most models), $\delta$ is the fractional overdensity of the patch, $d v_\parallel / d r_\parallel$ is the gradient of the proper velocity along the line of sight, $T_\gamma$ is the brightness temperature of the background light (usually the CMB), and $T_S$ is the spin temperature, or excitation temperature of the 21-cm transition. During the Dark Ages, the spin temperature is principally set by a competition between scattering of photons from the backlight (driving $T_S \rightarrow T_\gamma$) and collisions in the intergalactic medium (driving $T_S \rightarrow T_K$, the kinetic temperature of the gas). The black dashed curve in Figure~\ref{fig:sample-histories} follows the average spin temperature in the standard cosmological model (using \texttt{CosmoRec}; \cite{Chluba2011}). At $z > 150$, Compton scattering off of the residual free electrons in the IGM makes $T_K \approx T_\gamma \approx T_S$, so the 21-cm signal is extremely small. At $z \sim 150$, Compton scattering becomes inefficient and the gas cools relative to the CMB. Because the gas density is large, collisions couple the spin and kinetic temperatures. As the Universe expands and the density declines, collisions become more rare, and eventually the spin temperature returns to the CMB temperature. Only later, at $z < 25$ in this model, do the first luminous sources switch the 21-cm signal back on. It is this first absorption peak, at $150 > z > 30$, where the 21-cm probe offers a unique probe of cosmology. (The importance of observing  this sky-averaged absorption trough is discussed in \cite{Burns2019}; here we focus on the next step, measuring fluctuations in the background.)

\vspace{-12pt} \section{The Ultimate Frontier: Cosmology During the Dark Ages} \vspace{-12pt} 

\begin{SCfigure}
	\includegraphics[width=7.5cm]{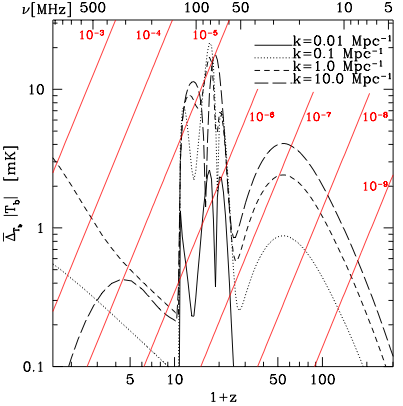} 
    \caption{{\bf 21-cm fluctuations are substantial during the Dark Ages.} The curves show the amplitude of the 21-cm brightness temperature fluctuations at several different wavenumbers from the Dark Ages to low redshifts, in a standard model of cosmology. Red diagonal lines compare these fluctuations to the foreground brightness temperature (from Galactic synchrotron): each scales the foreground by the number shown. Note that exotic cooling scenarios described in \S \ref{opp} could significantly amplify these fluctuations. From \cite{Pritchard:2012ja}.}
    \label{fig:pritchard-flucs}
\end{SCfigure}

Equation~(\ref{eq:21cm-dtb}) is the brightness temperature of a single patch in the IGM; during the Dark Ages, the density field --- and, along with it, the spin temperature and velocity fields --- vary over a wide range of scales.  Maps of 21-cm emission can therefore provide a sensitive probe of the power spectrum of density fluctuations. Figure~\ref{fig:pritchard-flucs} shows examples of how modes of the 21-cm fluctuation power spectrum evolve in the standard cosmology. 

The 21-cm power spectrum during the Dark Ages offers several advantages over other probes of the density field. Because they can use a spectral line to establish the distance to each patch, 21-cm measurements probe three-dimensional volumes --- unlike the CMB, which probes only a narrow spherical shell around recombination. Additionally, the 21-cm line does not suffer from Silk damping, which depresses the CMB fluctuations on relatively large scales, but probes the early Universe when small-scale fluctuations are still linear (unlike the galaxy power spectrum at later times). The number of independent modes accessible through this probe is therefore \cite{Loeb:2004js}
\begin{equation}
N_{\rm 21cm} \sim 8 \times 10^{11} \left( {k_{\rm max} \over 3 \ \mbox{Mpc}^{-1}} \right)^3 \left( {\Delta \nu \over \nu} \right) \left( {1+z \over 100} \right)^{-1/2}.
\label{eq:nmax}
\end{equation}
The choice of $k_{\rm max}$ --- the smallest physical scale to be probed --- is not obvious. The Jeans length during the Dark Ages corresponds to $k_{\rm max} \sim 1000\ \rm{Mpc}^{-1}$. Accessing these small-sale modes in three dimensions would require an enormous instrument, but equation~(\ref{eq:nmax}) shows that even a more modest effort provides a massive improvement over the information contained in all the measurable modes of the CMB, $N_{\rm CMB} \sim 10^7$. Moreover, because the density fluctuations are still small, these modes remain in the linear or mildly non-linear regime, allowing a straightforward interpretation of them in terms of the fundamental parameters of our Universe \cite{Lewis:2007}.  This treasure trove of statistical measurements will support a number of precision cosmological measurements, even within the ``standard" cosmology:

\begin{SCfigure}
	\includegraphics[width=9cm, clip=True, trim=0cm 0.8cm 0cm 0cm]{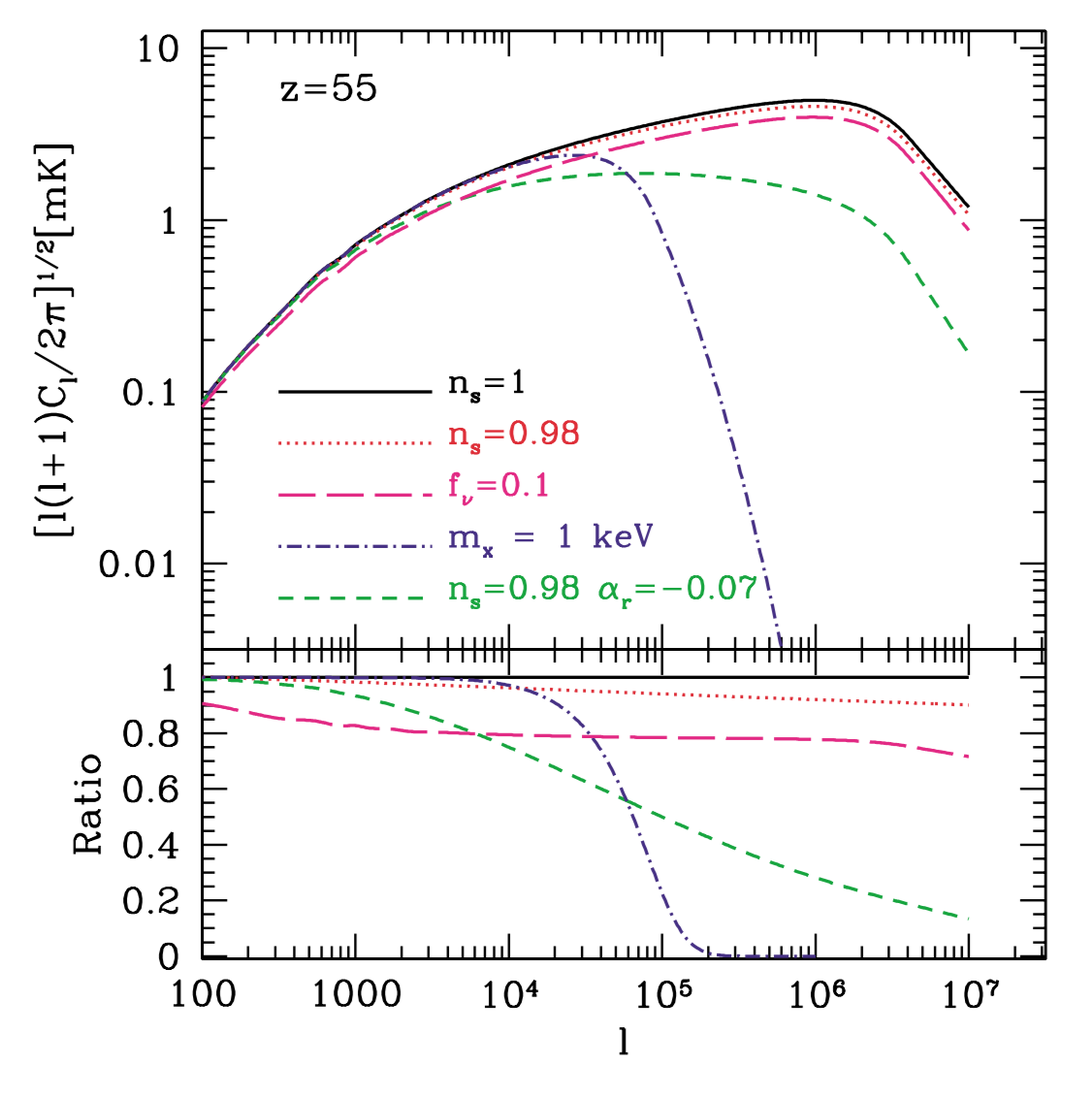} 
    \caption{{\bf The 21-cm power spectrum is a sensitive probe of cosmological parameters.} The angular power spectrum of 21-cm fluctuations at $z=55$ in ``standard" cosmologies, varying some of the parameters. The solid and dotted curves use $\Lambda$CDM with the power law index of density fluctuations $n_s=1$ and 0.98; respectively; the short-dashed curve adds a ``running" to the spectral index. The long-dashed curve assumes 10\% of the matter density is in the form of massive neutrinos (with masses 0.4~keV), while the dot-dashed curve assumes warm dark matter with particle masses of 1~keV.  From \cite{Loeb:2004js}.}
    \label{fig:sample-cos}
\end{SCfigure}

\begin{enumerate}

\vspace{-12pt} \item Because the 21-cm fluctuations extend to such small scales, they expand the dynamic range of power spectrum measurements over several orders of magnitude. Figure~\ref{fig:sample-cos} shows some examples of how cosmological parameters affect the power spectrum. The 21-cm measurements therefore can provide precision constraints on the running of the spectral index of the matter power spectrum \cite{Mao:2008jh}, a key parameter for inflation models for which stringent limits can test the underlying assumptions of the paradigm.

\vspace{-9pt} \item Another key constraint on the origin of density fluctuations will come from measurements of primordial non-Gaussianity imprinted by inflation. CMB measurements offer stringent constraints that can best be improved by larger-volume surveys --- but low-$z$ galaxy surveys suffer from contamination by non-Gaussianity developed through nonlinear structure formation. The clean, small-scale Dark Ages signal is an excellent opportunity to further constrain the existence of non-Gaussianity in the density power spectrum \cite{Chen2018} -- in principle, 21-cm Dark Ages measurements can place such tight limits that they can test the generic inflationary picture itself \cite{Munoz2015}. Other inflation probes are also possible (e.g., \cite{Shiraishi:2016jn}).

\vspace{-9pt} \item Several other cosmological parameters are best measured at small scales as well. For example, constraints on the total spatial curvature and neutrino masses improve by orders of magnitude \cite{Mao:2008jh}; the latter is also illustrated in Figure~\ref{fig:sample-cos}.

\vspace{-9pt} \item The extraordinary statistical precision of a 21-cm Dark Ages measurement would also allow us to identify deviations from a featureless primordial power spectrum that might be imprinted by complex models of inflation \cite{Chen:2016ft}.

\end{enumerate}

\vspace{-24pt} \section{Opportunities For the Next Decade: Exotic Physics in the Dark Ages} \label{opp} \vspace{-12pt} 

While the information contained in the Dark Ages 21-cm signal is clearly powerful, accessing the most useful of such measurements --- small-scale fluctuations over a broad redshift range --- will require a large radio telescope able to overcome the severe observational challenges \cite{Adshead:2011}, possibly on the lunar far side. But in the near term, pathfinders toward such an ultimate goal can constrain non-standard models of exotic particle physics models, including dark matter.

The black dashed curve in Fig.~\ref{fig:sample-histories} shows that the standard cosmology provides a clear and simple prediction for the spin temperature (and hence 21-cm power spectrum) throughout the Dark Ages. This era is a powerful probe of non-standard physics because the low gas temperature during this period makes the 21-cm line a sensitive calorimeter of additional heating (or cooling) and/or of an excess radio background over and above the CMB \cite{FengHolder2018}.

While many such processes tend to \emph{heat} the IGM and therefore \emph{decrease} the amplitude of the 21-cm signal, the recent evidence for a detection of a 21-cm absorption feature at 78~MHz by the EDGES collaboration \cite{Bowman2018} has triggered interest in non-standard models that \emph{amplify} the 21-cm signal (though note that the EDGES signal has not yet been independently confirmed). The EDGES feature is more than twice as deep than expected in a Universe that has been cooling adiabatically following the standard model. It therefore requires new physics that either \emph{cools} the IGM or \emph{increases} the radio background against which the gas absorbs. A panoply of new physics  has been proposed, including dark matter-baryon scattering \cite{MunozKovetz2015,Barkana2018,SlatyerWu2018,HiranoBromm2018}, millicharged dark matter \cite{MunozLoeb2018,Berlin2018,Kovetz2018}, dark matter annihilation \cite{cheung2019},  axions \cite{Moroi2018}, neutrino decay \cite{Chianese2019}, charge sequestration \cite{Falkowski2018}, quark nuggets \cite{Lawson2018}, dark photons \cite{Jia2019}, and interacting dark energy \cite{Costa2018}.

The same new physics may very well affect the Dark Ages signal, because it too depends strongly on the thermal evolution of the IGM. The solid curves in Figure~\ref{fig:sample-histories} illustrate how models that invoke excess cooling to explain the EDGES result (shown schematically by the gray contours) could also greatly amplify the Dark Ages signal. (Here the curves use a phenomenological parameterized cooling model as in \cite{Mirocha2019}; physically-motivated models will differ in the details.) Degeneracies between astrophysics and exotic physics in $z<20$ measurements could thus be broken by observations of the Dark Ages, which are not affected by astrophysical sources.

\begin{SCfigure}
	\includegraphics[width=9cm, clip=True, trim=0cm 0.6cm 0cm 0.5cm]{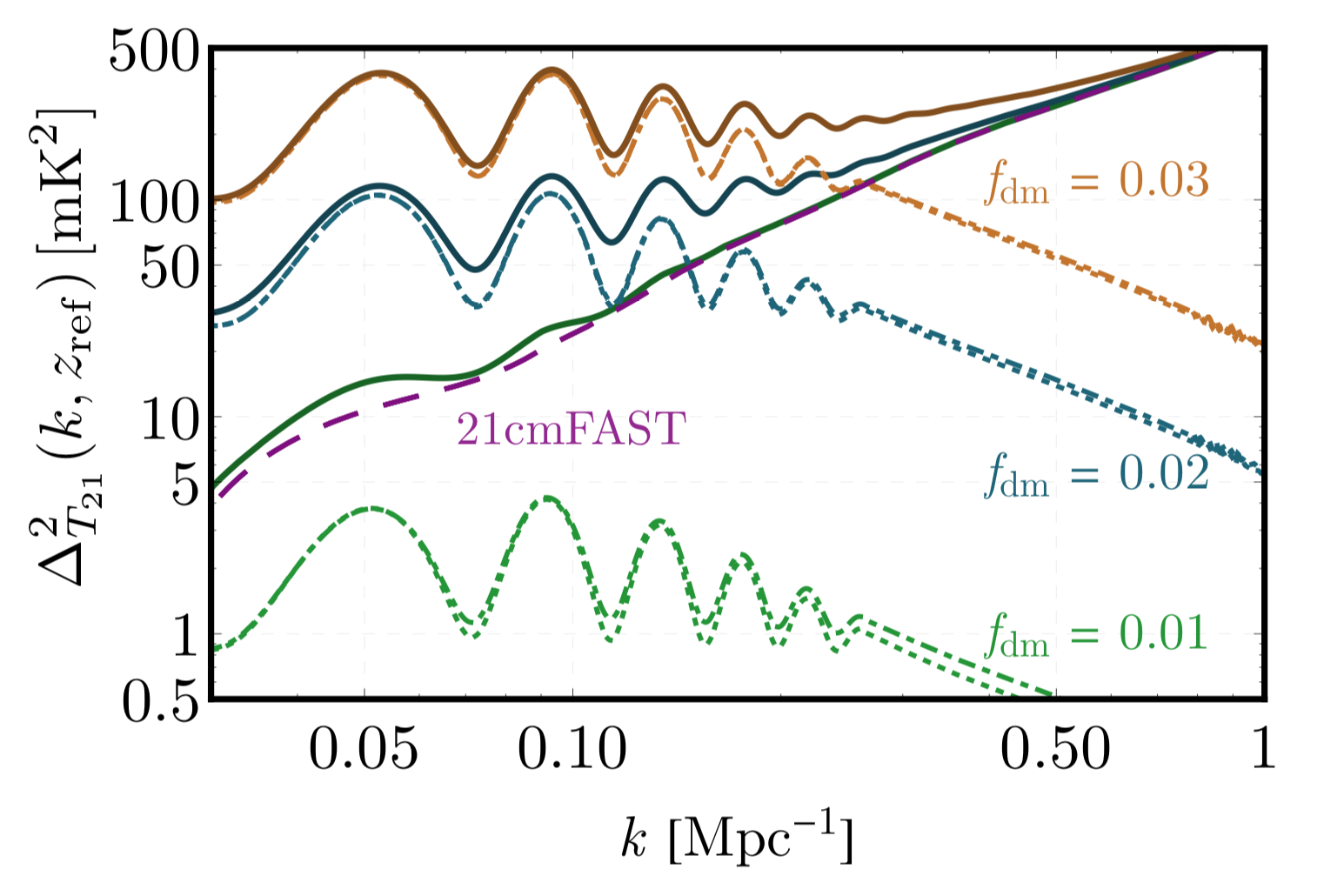} 
    \caption{{\bf Models that invoke exotic physics imprint signatures on the 21-cm power spectrum.} In these models, a fraction $f_{\rm dm}$ of the dark matter is assumed to have a small charge; the oscillations in the power spectrum arise from the large-scale streaming of baryons relative to dark matter. The solid curves are the total power, while the others show the contribution from dark matter-baryon scattering. 
    From \cite{Munoz2018}.}
    \label{fig:munoz-dm}
\end{SCfigure}

Moreover, any model producing either excess IGM cooling or an excess radio background will inevitably affect the 21-cm power spectrum. For example, if the cooling is triggered by scattering between a fraction of the dark matter that has a modest charge, the scattering rate will be modulated by the relative velocity of dark matter and baryons, which has large-scale structure imprinted before recombination. Such a model therefore leaves distinct features in the 21-cm power spectrum, as shown in Fig.~\ref{fig:munoz-dm} \cite{Munoz2018}. The detailed implications of most of these exotic models are mostly unexplored, but any model in which (1) the energy exchange depends on local density, velocity, or temperature; or (2) in which background radio energy deposition is inhomogeneous should generically leave signatures in the power spectrum. 

Although most of these exotic cooling and/or radio background models were inspired by the EDGES measurement, constraints on such scenarios are interesting even if that result is not confirmed by future measurements --- given the apparent difficulties in detecting dark matter directly and in producing it in accelerators, it appears increasingly likely that dark matter is not simple after all. Constraints on exotic models and interactions will therefore be essential for pinning down its properties. While many of these scenarios will affect the 21-cm signal at later times (as described in the white papers by Mirocha et al. and Liu et al.), the Dark Ages still offer the ``cleanest" probe of the underlying physics.

\vspace{-12pt} \section{Conclusion} \vspace{-12pt}

Measurements of the 21-cm line power spectrum from the Dark Ages offer extraordinary promise for fundamental cosmology.  Many observational challenges need to be solved before such an experiment can become feasible, including those related to the terrestrial environment (the ionosphere and man-made or atmospheric interference --- which can be avoided with a lunar instrument), but also including foreground removal and calibration strategies.  Ground-based experiments are currently addressing many of these issues while targeting a first detection of the 21-cm signal from the Epoch of Reionization.  Assuming these problems can be addressed, calculations following \cite{Pober:2016} suggest that a first detection of the 21-cm power spectrum from the Dark Ages (in the standard cosmology) will require a filled aperture radio interferometer with a collecting area of $\sim 5$ square kilometers.  Extracting precision cosmological constraints will require an even larger array.  However, sensitivity scales approximately linearly with collecting area; therefore, the enhanced signals discussed in \S\ref{opp} could be detectable with instruments similar to existing terrestrial experiments.  {\bf Over the next decade, efforts to detect exotic new physics with the Dark Ages 21-cm fluctuation power spectrum may be within reach, while also serving as pathfinder experiments for the ultimate frontier of cosmology.}

\pagebreak

\bibliographystyle{abbrv}
\bibliography{dark-ages-21cm} 

\end{document}